\documentclass[]{aa}       % [onecolumn,referee]
\usepackage{graphicx}
\usepackage{txfonts}
\usepackage{bm}

\begin{document}
\def\bea{\begin{eqnarray}}
\def\eea{\end{eqnarray}}
\def\be{\begin{equation}}
\def\ee{\end{equation}}
\def\rra{\right\rangle}
\def\lla{\left\langle}
\def\le#1{\label{eq:#1}}
\def\re#1{\ref{eq:#1}}
\def\eps{\epsilon}
\def\sgm{\Sigma^-}
\def\la{\Lambda}
\def\kv{\bm{k}}
\def\vd{{\cal V}} 
\def\pp{{\cal P}} 
\def\kk{{\cal K}} 

\title{Protoneutron stars within the Brueckner-Bethe-Goldstone theory}

\author{O. E. Nicotra, M. Baldo, G. F. Burgio, and H.-J. Schulze}

\institute{Istituto Nazionale di Fisica Nucleare, Sezione di Catania, 
and Dipartimento di Fisica, Universit\'a di Catania,
Via S. Sofia 64, I-95123 Catania, Italy}

% \noindent
\date{Received / Accepted}

% \bigskip

\abstract{
We study the structure of newly born neutron stars (protoneutron stars) 
within the finite temperature Brueckner-Bethe-Goldstone theoretical approach
including also hyperons.
We find that for purely nucleonic stars both finite temperature and 
neutrino trapping reduce the value of the maximum mass. 
For hyperonic stars the effect is reversed,
because neutrino trapping
shifts the appearance of hyperons to larger baryon density
and stiffens considerably the equation of state.
%\vskip20mm
}

%\authorrunning{O. E. Nicotra et al.} 
%\titlerunning{Protoneutron stars within the BBG theory}

%\pacs{ 
% 21.65.+f,  % Nuclear matter
% 26.60.+c,  % Nuclear aspects of neutron stars
% 24.10.Cn,  % Many-body theory 
% 97.60.Jd,  % Neutron stars
%     }

\maketitle

%--------|---------|---------|---------|---------|---------|---------|---------|
\section{Introduction}

A protoneutron star (PNS) is formed after a successful supernova explosion 
%as the stellar remnant becomes gravitationally decoupled from the expanding ejecta 
and constitutes for several tens of seconds a transitional state 
to either a neutron star or a black hole
(Prakash et al. 1997). 
Initially, the PNS is optically thick to neutrinos, that is, they
are temporarily trapped within the star. 
The subsequent evolution of the PNS is dominated by neutrino diffusion,
which first results in deleptonization and subsequently in cooling. 
After a much longer time, photon emission competes with neutrino
emission in neutron star cooling.

In this paper, we will focus upon the essential ingredient that
governs the macrophysical evolution of neutron stars, i.e., 
the equation of state (EOS) of dense matter at finite temperature. 
We have developed a microscopic EOS
in the framework of the Brueckner-Bethe-Goldstone (BBG) many-body approach 
extended to finite temperature. 
This EOS has been successfully applied to the study 
of the limiting temperature in nuclei (Baldo et al. 1999, 2004). 
The scope of this work is to present results on the composition and structure 
of these newly born stars with the EOS previously mentioned.

Two new effects have to be considered for a PNS. 
First, thermal effects which result in entropy production, with values
of few units per baryon, and temperatures up to 30-40 MeV
(Burrows \& Lattimer 1986; Pons et al. 1999). 
Second, the fact that neutrinos are trapped in the star, which means that 
the neutrino chemical potential is non-zero.
This alters the chemical equilibrium and leads to compositional changes. 
Both effects may result in observable consequences in the neutrino signature 
from a supernova and may also play an important role in determining whether 
or not a given supernova ultimately produces a cold neutron star or a 
black hole. 

This paper is organized as follows. 
In Section 2 we briefly illustrate the BBG
many-body theory at finite temperature. 
Section 3 is devoted to the study of the 
stellar matter composition, both with and without neutrino trapping, 
and the resulting EOS. 
In Section 4 we discuss our results regarding the structure of 
(proto)neutron stars, in particular their maximum mass. 
Finally, conclusions are drawn in Section 5.

%------------------------------------------------------------------------------
\section{Brueckner theory}

Over the last two decades the increasing interest for the 
EOS of nuclear matter has stimulated a great deal of
theoretical activity. 
Phenomenological and microscopic models of
the EOS have been developed along parallel lines with
complementary roles. 
The former models include nonrelativistic
mean field theory based on Skyrme interactions (Bonche et al. 1998) and
relativistic mean field theory based on meson-exchange
interactions (Walecka model) (Serot \& Walecka 1986). 
Both of them fit the parameters of the interaction in order 
to reproduce the empirical saturation properties of nuclear matter 
extracted from the nuclear mass table. 
The latter ones include nonrelativistic
Brueckner-Hartree-Fock (BHF) theory (Baldo 1999) and its relativistic
counterpart, the Dirac-Brueckner theory (Machleidt 1989; Li et al. 1992), 
the nonrelativistic variational approach also corrected by
relativistic effects (Akmal et al. 1997, 1998; Morales et al. 2002), 
and more recently the chiral perturbation theory
(Kaiser et al. 2002). 
In these approaches the parameters
of the interaction are fixed by the experimental nucleon-nucleon
and/or nucleon-meson scattering data. 

For states of nuclear matter
with high density and high isospin asymmetry the experimental
constraints on the EOS are rather scarse and indirect. 
Different approaches lead to different or even contradictory
theoretical predictions for the nuclear matter properties.
The interest for these properties lies, to a large extent, 
in the study of dense astrophysical objects, i.e.,
supernovae and neutron stars. 

One of the most advanced microscopic approaches to the EOS of 
nuclear matter is the Brueckner theory. 
In the recent years, it has made a rapid
progress in several aspects:
(i) The convergence of the
Brueckner-Bethe-Goldstone expansion has been firmly
established (Song et al. 1998; Baldo et al. 2000b). 
(ii) The addition of phenomenological three-body forces (TBF) based on 
the Urbana model 
(Carlson et al. 1983; Schiavilla et al. 1986),
permitted to improve to a large extent the agreement with the empirical
saturation properties
(Baldo et al. 1997; Zhou et al. 2004).
(iii) The BHF approach has been extended in a fully microscopic and 
self-consistent way to describe nuclear matter containing also 
hyperons (Schulze et al. 1995, 1998), 
opening new fields of applications such as hypernuclei 
(Cugnon et al. 2000; Vida\~na et al. 2001)
and a more realistic modeling of neutron stars (Baldo et al. 1998, 2000a).

In this paper, we will discuss the nuclear EOS at 
finite temperature within the BBG theory, 
and apply it to the study of protoneutron stars.
For this purpose, we have performed a systematic calculation of the EOS within 
the Bloch-De Dominicis formalism with a realistic nucleon-nucleon 
interaction, including nucleonic three-body forces.
Details are given in the following section.

\subsection{EOS of nuclear matter at finite temperature}

Only few microscopic calculations of the nuclear EOS at finite temperature
are so far available. 
The variational calculation by Friedman \& Pandharipande (1981) was one of 
the first semi-microscopic investigations of the finite temperature EOS. 
The results predict a Van der Waals behavior, which leads to a liquid-gas 
phase transition with a critical temperature \hbox{$T_c \approx$ 18--20 MeV.} 
Later, Brueckner calculations (Lejeune et al. 1986; Baldo \& Ferreira 1999) 
and chiral perturbation theory at finite temperature 
(Kaiser et al. 2002) 
confirmed these findings with very similar values of $T_c$.
The Van der Waals behavior was also found in the finite temperature
relativistic Dirac-Brueckner calculations of 
(Ter Haar \& Malfliet 1986, 1987; Huber et al. 1999),
although at a lower temperature.

The formalism which is closest to the BBG expansion, 
and actually reduces to it in the zero temperature limit, 
is the one formulated by Bloch \& De Dominicis (1958, 1959a, 1959b).
In this approach the essential ingredient is the two-body scattering 
matrix $K$, which, along with the single-particle potential $U$, 
satisfies the self-consistent equations
\bea
  \lla k_1 k_2 | K(W) | k_3 k_4 \rra
 &=& \lla k_1 k_2 | V | k_3 k_4 \rra  
\nonumber\\&&
 \hskip -27mm +\; \mathrm{Re} \sum_{k_3' k_4'} 
 \langle k_1 k_2 | V | k_3' k_4' \rangle 
 { [1-n(k_3')] [1-n(k_4')] \over 
   W - E_{k_3'} - E_{k_4'} + i\eps }
%\nonumber\\&& \hskip-7mm\times
 \langle k_3' k_4' | K(W) | k_3 k_4 \rangle
\nonumber\\&&
\label{eq:kkk}
\eea
and
\be
 U(k_1) = \sum_{k_2} n(k_2) \lla k_1 k_2 | K(W) | k_1 k_2 \rra_A \:,
\label{eq:ueq}
\ee
where $k_i$ generally denote momentum, spin, and isospin.
Here $V$ is the two-body interaction, 
$W = E_{k_1} + E_{k_2}$ represents the starting energy, and
$E_k = k^2/2m + U(k)$ the single-particle energy.
Eq.~(\ref{eq:kkk}) coincides with the Brueckner equation for the $K$ matrix
at zero temperature, 
if the single-particle occupation numbers $n(k)$ are taken at $T = 0$. 
At finite temperature $n(k)$ is a Fermi distribution.
For a given density and temperature, 
Eqs.~(\ref{eq:kkk}) and (\ref{eq:ueq}) have to be solved 
self-consistently along with the following equation for the auxiliary
chemical potential $\tilde{\mu}$,
\be
 \rho = \sum_k n(k) = 
 \sum_k {1\over e^{\beta (E_k - \tilde{\mu})} + 1 } \:.
\label{eq:ro}
\ee
The grand-canonical potential density $\omega$ can be written 
(Baldo \& Ferreira 1999), 
\bea
 \omega &=& \omega_0 - \sum_k n(k)U(k) 
\nonumber\\
 && \hskip5mm +\; {1\over 2}\sum_{k_1,k_2} n(k_1)n(k_2) 
      \lla k_1 k_2 | K(W) | k_1 k_2 \rra_A \:,
\label{eq:om}
\eea
where $\omega_0$ is the grand-canonical potential density for a gas 
of independent particles with single-particle spectrum $E_k$,
\be
  \omega_0 = -{1\over \beta} \sum_k 
  \ln\left( 1 +  e^{-\beta (E_k - \tilde{\mu})} \right) \:.
\ee 
Eq.~(\re{om}) neglects a series of terms proportional to
{$n(k)[1-n(k)]$} (or powers of it), 
which turn out to be negligible in the
temperature and density ranges relevant for neutron and protoneutron
stars (Baldo 1999). 
The free energy density is then
\be
 f = \omega + \rho \tilde{\mu} 
\ee
and the ``true" chemical potential $\mu$ and the pressure $p$ are given by
\bea
 \mu &=& {{\partial f}\over{\partial \rho}} \:,
\\
 p &=& \rho^2 {{\partial (f/\rho)}\over{\partial \rho} } \:.
\eea  
We stress that this procedure permits to fulfill the Hugenholtz-Van Hove 
theorem in the calculation of thermodynamical quantities in the Brueckner 
theory.
For an extensive discussion of this topic, the reader is referred 
to (Baldo 1999), and references therein.

\begin{figure}[t] %............................................................
\centering
\includegraphics[height=11cm,angle=0,bb=0 50 500 730,clip]{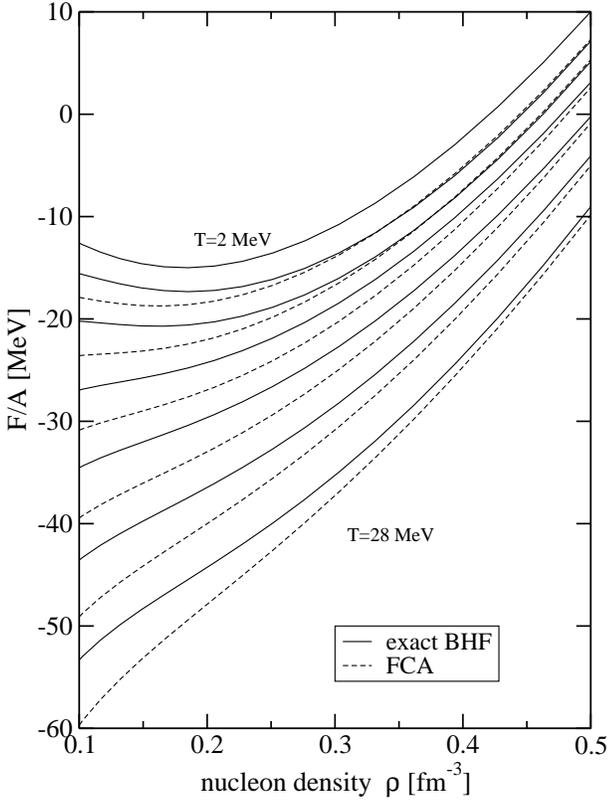}
\caption{
Free energy per particle in exact BHF calculation (solid lines)  
and Frozen Correlations Approximation (dashed lines).
The different curves correspond to temperatures  
$T=2$,8,12,16,20,24,28 MeV from top to bottom.}
\label{f:fa}
\end{figure}

The determination of the different configurations inside protoneutron stars
requires the calculation of the EOS at different chemical compositions
in search of the beta equilibrium. 
To save computational time and simplify
the numerical work we introduce an approximate procedure. 
For each configuration the single-particle potential $U_i(k)$
for the component $i$ is calculated self-consistently at $T = 0$.
At $T \neq 0$ we consider $U_i(k)$ independent of $T$. 
In this approximation the correlations at $T \neq 0$ 
are assumed to be essentially the same as at $T = 0$ 
({\it Frozen Correlations Approximation}).
It turns out that the assumed independence is 
valid to a good accuracy (Baldo \& Ferreira 1999, Fig.~12), 
at least for not too high temperature, 
due to a substantial compensation between the 
$T$-dependence of the $K$-matrix and the $T$-dependence of the Fermi 
distributions, see Eq.~(\re{ueq}). 
Once this approximation is introduced,
the grand-canonical potential density $\omega$ can be trivially calculated
and the free energy density has the following simplified expression
\be
 f = \sum_i \left[ \sum_{k} n_i(k)  
 \left( {k^2\over 2m_i} + {1\over 2}U_i(k) \right) - Ts_i \right] \:,
\ee
where 
\be
 s_i = - \sum_{k} \Big( n_i(k) \ln n_i(k) + [1-n_i(k)] \ln [1-n_i(k)] \Big)  
\ee
is the entropy density for component $i$ 
treated as a free gas with spectrum $E_i(k)$. 
To illustrate the accuracy of this approximation, in Fig.~\ref{f:fa}
is reported the comparison between the Frozen Correlations Approximation 
and the full microscopic calculations for the free energy in symmetric 
nuclear matter in the relevant density range. 
For pure neutron matter the agreement is even closer.

 %.................................................................

\begin{figure*}[t] %...........................................................
\centering
\includegraphics[height=17cm,angle=270,bb=50 30 560 700,clip]{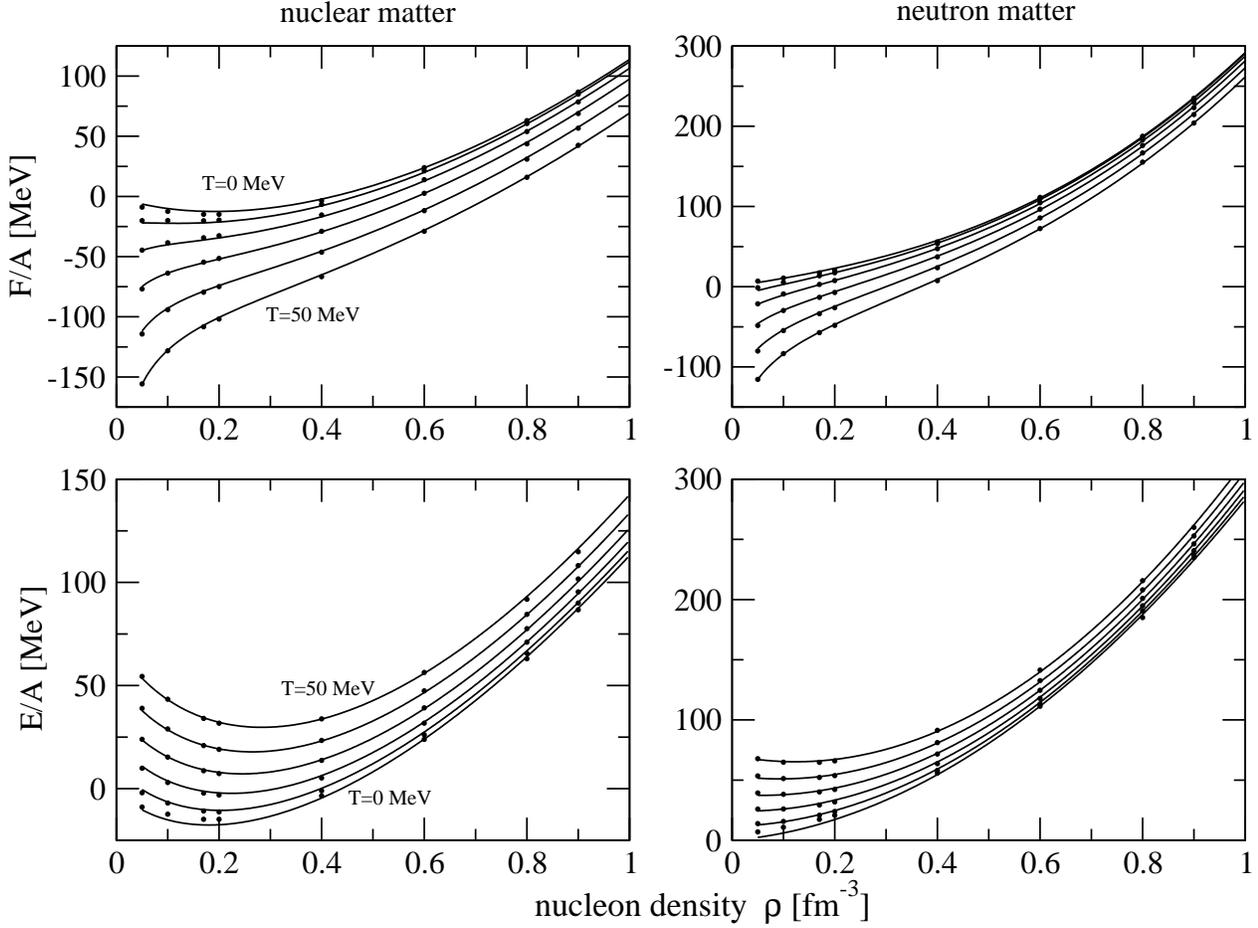}
\caption{
Finite temperature BBG equation of state for symmetric (left-hand panels) and 
purely neutron (right-hand panels) matter. 
The upper panels show the free energy and the lower panels the 
internal energy per particle as a function of the nucleon density.
The temperatures vary from 0 to 50 MeV in steps of 10 MeV.}
\label{f:ba}
\end{figure*} %................................................................

\subsection{Effects of three-body forces}

It is well known that at zero temperature the non-relativistic 
microscopic approaches do not reproduce correctly the nuclear matter 
saturation point
$\rho_0 \approx 0.17~\mathrm{fm}^{-3}$, $E/A \approx -16$ MeV. 
One common way of correcting this deficiency 
is to introduce three-body forces among nucleons. 
A complete microscopic theory of
TBF is not available yet, thus one is forced to use phenomenological models.
We have adopted the so-called Urbana model 
(Carlson et al. 1983; Schiavilla et al. 1986),
which consists of an attractive term due to two-pion exchange with 
excitation of an intermediate $\Delta$ resonance, 
and a repulsive phenomenological central term.
In the BBG approach the TBF is reduced to a density dependent two-body force 
by averaging over the position of the third particle, 
assuming that the probability of having two particles at a
given distance is reduced according to the two-body correlation function.
The corresponding EOS satisfies several requirements, namely, 
(i) it reproduces correctly the nuclear matter saturation point 
(Baldo et al. 1997; Zhou et al. 2004), 
(ii) the incompressibility is compatible with values extracted from 
phenomenology (Myers \& Swiatecki 1996), 
(iii) the symmetry energy is compatible with nuclear phenomenology, 
(iv) the causality condition is fulfilled up to the densities 
typical of neutron star cores.
In all calculations presented in this paper we use the Argonne $V_{18}$ 
nucleon-nucleon potential (Wiringa et al. 1995) 
along with the phenomenological Urbana TBF.

In Fig.~\ref{f:ba} we display the EOS obtained following the above discussed 
procedure, both for symmetric and purely neutron matter. 
In the upper panels we display the free energy, 
whereas the lower panels show the internal energy per particle as a function of 
the nucleon density, for several values of the temperature
between 0 and 50 MeV. 
%The dots represent the calculated values and the solid lines the spline fitting. 
We notice that the free energy of symmetric matter shows a typical 
Van der Waals behavior (with $T_C\approx 16$ MeV)
and is a monotonically decreasing function of the temperature.  
At $T=0$ the free energy coincides with the total energy and the corresponding 
curve is just the usual nuclear matter saturation curve. 
On the contrary, the internal energy
is an increasing function of the temperature. 
The effect is less pronounced for pure neutron matter
due to the larger Fermi energy of the neutrons at given density.

\subsection{Inclusion of hyperons}

It is commonly accepted that, 
whereas at moderate densities $\rho \approx \rho_0$ the matter inside 
a neutron star consists only of nucleons and leptons, 
at higher densities other baryonic species may appear due to the fast rise of 
the baryon chemical potentials with density (Glendenning 1982, 1985). 
For this purpose we have extended the BBG approach in order to include the 
$\Sigma$ and $\la$ hyperons (Schulze et al. 1998; Baldo et al. 1998, 2000a).
The inclusion of hyperons is a delicate task, and requires the knowledge of the
nucleon-hyperon (NH) and hyperon-hyperon (HH) interactions. 
In our work we have used the Nijmegen soft-core NH potential 
(Maessen et al. 1989),
that is well adapted to the available experimental NH scattering data.
Unfortunately, no HH scattering data 
and therefore no reliable HH potentials are available at the moment.
We have therefore neglected these interactions in our calculations,
which is supposedly justified, as long as the hyperonic partial 
densities remain limited.

We have found that due to its negative charge the $\sgm$ hyperon is the 
first strange baryon appearing in the reaction $n+n \rightarrow p+\sgm$,
in spite of its substantially larger mass compared to the neutral 
$\la$ hyperon ($M_{\sgm}=1197\;\mathrm{MeV}, M_\la=1116\;\mathrm{MeV}$).
The presence of hyperons strongly softens the EOS,
mainly due to the larger number of baryonic degrees of freedom.
This EOS produces a maximum neutron star mass that lies
slightly below the canonical value of 1.44 $M_\odot$ (Taylor \& Weisberg 1989),
which could indicate the presence of non-baryonic (quark) matter 
in the interior of heavy neutron stars
(Burgio et al. 2002a, 2002b; Baldo et al. 2003, Maieron et al. 2004). 
However, the quantitative effects of more reliable NH and HH potentials
as well as hyperonic TBF need still to be explored in the future.

For these reasons, 
we will present in this article only zero-temperature results with interacting
hyperons and perform finite-temperature calculations 
in a much simpler way
using non-interacting hyperons
in order to estimate the sensitivity of the main global observables
to finite temperature.

\begin{figure*}[t] %...........................................................
\centering
\includegraphics[height=17cm,angle=270,bb=70 50 570 740,clip]{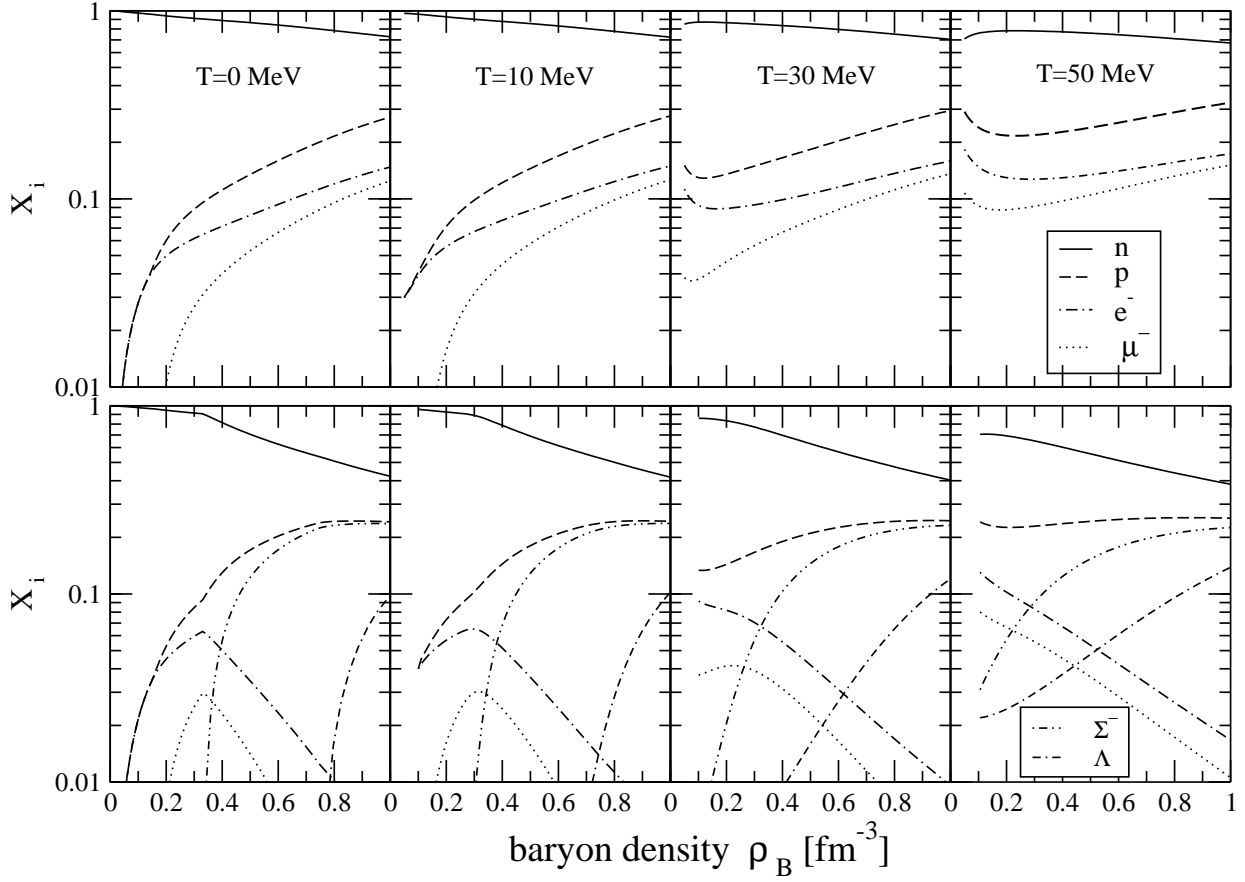}
\caption{
Relative populations for neutrino-free matter as a function of 
the baryon density for several values of the temperature. 
Upper panels show the partial concentrations for the purely nucleonic case, 
whereas in the lower panels noninteracting hyperons 
$\sgm$ and $\la$ are included as well.}
\label{f:xi}
\end{figure*} %................................................................

\begin{figure*}[t] %...........................................................
\centering
\includegraphics[height=17cm,angle=270,bb=70 30 580 740,clip]{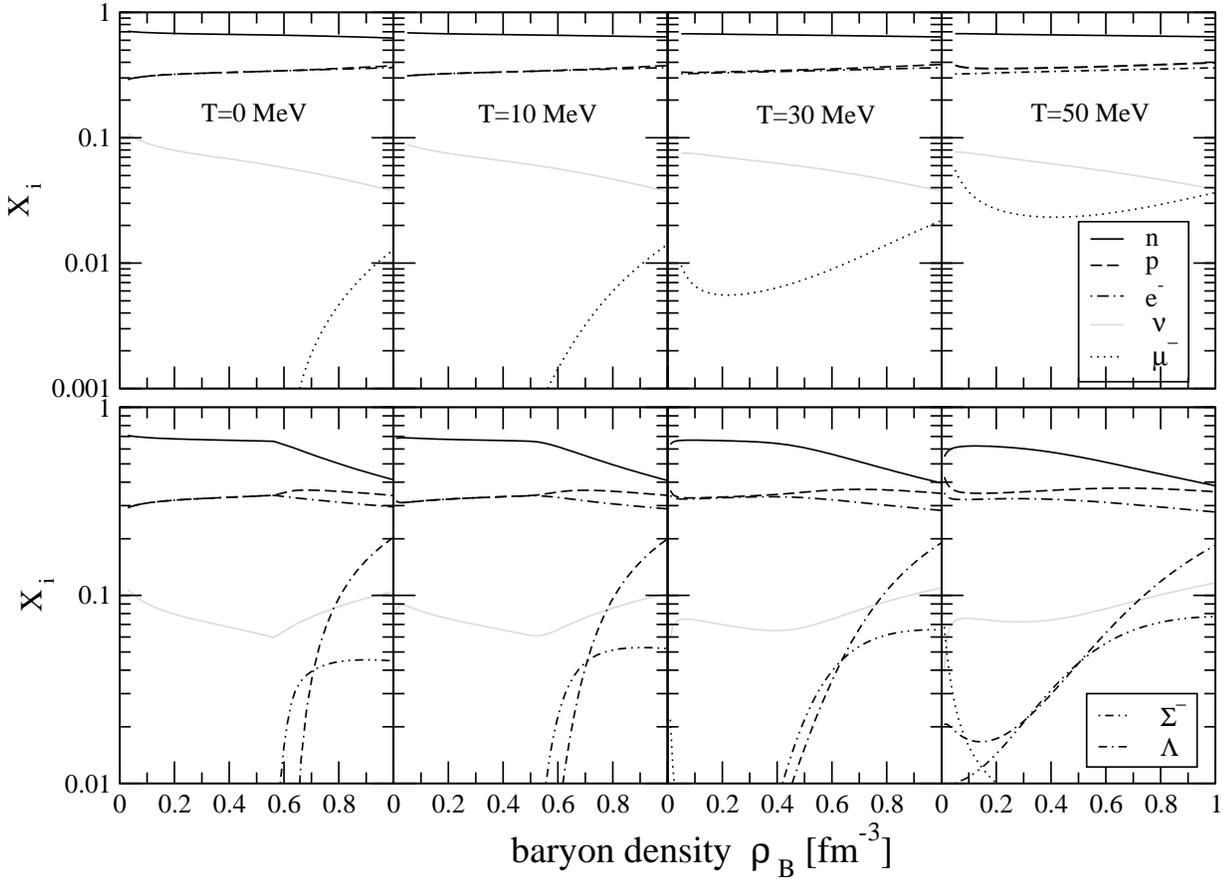}
\caption{
Same as Fig.~\ref{f:xi}, but for neutrino-trapped matter.}
\label{f:xinu}
\end{figure*} %................................................................

\begin{figure}[t] %............................................................
\includegraphics[width=9.5cm,angle=270,bb=90 130 560 130]{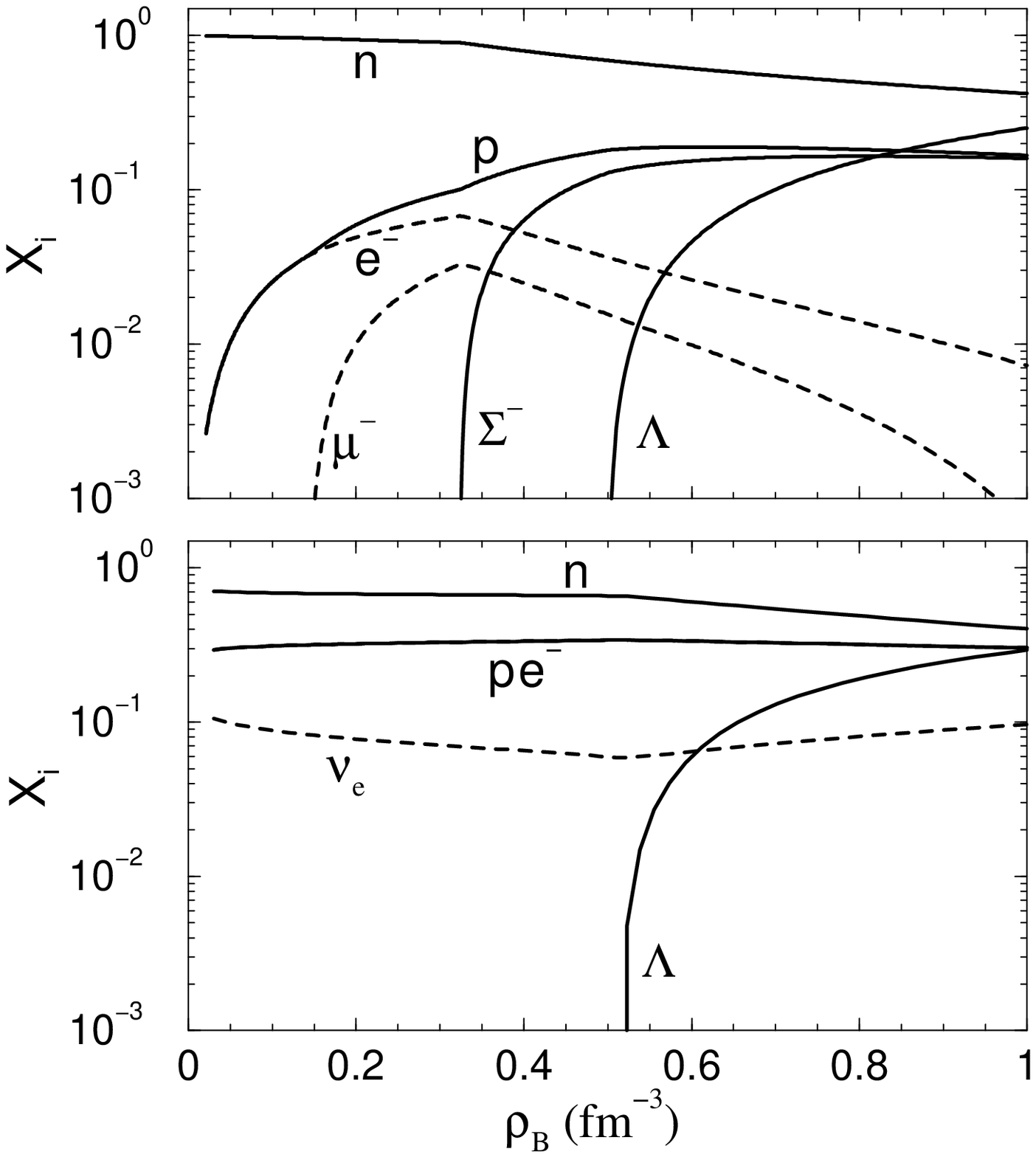}
\caption{
Composition of cold beta-stable matter with interacting hyperons, 
for neutrino-free (upper panel) and neutrino-trapped (lower panel) matter.}
\label{f:xiint}
\end{figure} %.................................................................

%------------------------------------------------------------------------------
\section{Composition and EOS of hot stellar matter}

For stars in which the strongly interacting particles are only baryons, 
the composition is determined by the requirements of charge 
neutrality and equilibrium under the weak processes
\be
 B_1 \rightarrow B_2 + l + {\overline \nu}_l \ ,\quad   
 B_2 + l \rightarrow B_1 + \nu_l \:,
\label{weak:eps}
\ee
where $B_1$ and $B_2$ are baryons and $l$ is a lepton, 
either an electron or a muon. 
When the neutrinos have left the system, these two requirements imply that 
the relations
\be
 \sum_{i}q_{i}x_{i}+\sum_{l}q_{l}x_{l}=0
\label{neutral:eps}
\ee
and
\be
 \mu_{i} = b_{i}\mu_{n}-q_{i}\mu_{l}
\label{mufre:eps}
\ee
are satisfied. 
In the expression above, $x_i=\rho_i/\rho_B$ represents the 
baryon fraction for the species $i$, and $\rho_B$ the baryon density.
The neutron chemical potential is denoted by $\mu_n$, whereas $\mu_i$  
refers to the chemical potential of the baryon species $i$, $b_{i}$ to its 
baryon number and $q_{i}$ to the electric charge. 
The same holds true for the quantities with subscript $l$. 
Under the condition that the neutrinos are trapped in the 
system, the beta equilibrium equation (\ref{mufre:eps}) is altered to
\be
 \mu_i = b_i \mu_n - q_i( \mu_l - \mu_{\nu_l} ) \:,
\label{nitrap:eps}
\ee
where $\mu_{\nu_l}$ is the chemical potential of the neutrino $\nu_{l}$.

For stellar matter containing nucleons and hyperons as relevant degrees of 
freedom, the chemical equilibrium conditions read explicitly
\bea
 \mu_n - \mu_p &=& \mu_e - \mu_{\nu_e} = \mu_\mu + \mu_{\bar{\nu}_\mu} \:, 
\nonumber\\  
 \mu_{\sgm} &=& 2 \mu_n - \mu_p \:, 
\nonumber\\ 
  \mu_\la  &=& \mu_n \:.
\label{beta:eps}
\eea
Because of trapping, the numbers of leptons per baryon 
of each flavor $l=e,\mu$,
\be
 Y_l = x_l + x_{\nu_l} - x_{\bar{\nu}_l} \:,
\label{lepfrac:eps}
\ee
are conserved on dynamical time scales. 
Gravitational collapse 
calculations of the white-dwarf core of massive stars indicate that at 
the onset of trapping, the electron lepton number 
$Y_e = x_e + x_{\nu_e}\simeq 0.4$, 
the precise value depending on the efficiency of electron capture 
reactions during the initial collapse stage. 
Also, because no muons are present when neutrinos become trapped, 
the constraint 
$Y_\mu = x_\mu - x_{\bar{\nu}_\mu} = 0$ 
can be imposed. 
We fix the $Y_l$ at these values in our calculations 
for neutrino-trapped matter.

The determination of the chemical potentials is one of the most 
important and delicate steps of the entire calculation. 
In fact, in asymmetric nuclear matter,
it involves the knowledge of the free energy and its partial 
derivatives with respect to the total baryon density and proton fraction, i.e.,
\bea
 \mu_{n}(\rho,x_{p}) &=& 
 \left[ 1 + \rho \frac{\partial }{\partial\rho}
        -x_{p}\frac{\partial }{\partial x_{p}} \right] {f\over\rho} \:,
\label{mun:eps}
\\
 \mu_{p}(\rho,x_{p}) &=& 
  \left[ 1 + \rho\frac{\partial}{\partial\rho} 
         + (1-x_p)\frac{\partial}{\partial x_p} \right] {f\over\rho} \:.
\label{mup:eps}
\eea
The derivatives are computed by using an analytical fit of the free 
energy $f(T,\rho,x_{p})$, 
which is given in Appendix~A.

As far as the hyperon chemical potentials are concerned, 
they are obtained through the 
standard procedure at zero temperature illustrated in (Maieron et al. 2004).
The chemical potentials of the noninteracting leptons and hyperons 
at finite temperature are obtained by solving 
numerically the free Fermi gas model at finite temperature. 

Once the baryonic and leptonic chemical potentials are known, 
one can proceed to calculate the composition of the $\beta$-stable 
stellar matter,
and then the free energy density $f$ and pressure $p$ 
through the usual thermodynamical relation
\be
 p = \rho^2 {\partial{(f/\rho)}\over \partial{\rho}} \:.
\ee
The stable configurations of a (proto)neutron star can be obtained from the 
well-known hydrostatic equilibrium equations of Tolman, Oppenheimer, and Volkov
(Shapiro \& Teukolsky 1983)
for the pressure $p$ and the enclosed mass $m$
\bea
 {dp(r)\over dr} &=& -\frac{Gm(r)\eps(r)}{r^2}
 \frac{\big[ 1 + {p(r)\over\eps(r)} \big]
       \big[ 1 + {4\pi r^3p(r)\over m(r)} \big]}
 {1-{2Gm(r)\over r}} \:,
\label{tov1:eps}
\\
 \frac{dm(r)}{dr} &=& 4\pi r^{2}\eps(r) \:,
\label{tov2:eps}
\eea
once the EOS $p(\eps)$ is specified, being $\eps$
the total energy density ($G$ is the gravitational constant).
For a chosen central value of the energy density, the numerical integration of 
Eqs.~(\ref{tov1:eps}) and (\ref{tov2:eps}) provides the mass-radius relation.

For the description of the NS crust at zero temperature, 
we have joined the hadronic equations of state above described with the 
ones by Negele \& Vautherin (1973) in the medium-density regime 
($0.001\;\mathrm{fm}^{-3}<\rho<0.08\;\mathrm{fm}^{-3}$), 
and the ones by Feynman, Metropolis, \& Teller (1949) 
and Baym, Pethick, \& Sutherland (1971)
for the outer crust ($\rho<0.001\;\mathrm{fm}^{-3}$). 

Simulations of supernovae explosions 
(Burrows \& Lattimer 1986; Pons 1999) show that the
protoneutron star has neither an isentropic nor an isothermal profile.
For simplicity we will assume a constant temperature inside the star and
attach a cold crust for the outer part. 
This schematizes the temperature profile of the protoneutron star.

%------------------------------------------------------------------------------
\section{Results and discussion}

\subsection{Composition of stellar matter}

First we will discuss the populations of beta-equilibrated stellar matter, 
by solving the chemical equilibrium conditions given by Eqs.~(\ref{beta:eps}),
supplemented by electrical charge neutrality and baryon number conservation. 
In Fig.~\ref{f:xi} we display the relative particle fractions as a function 
of the baryon density for several values of the temperature 
$T=$ 0, 10, 30, and 50 MeV. 
The upper panels show the particle fractions when stellar matter contains 
only neutrons, protons, electrons, and muons,
whereas the lower panels include the appearance of $\sgm$ and $\la$ hyperons
treated as noninteracting particles.

First let us discuss the purely nucleonic case.
We notice that temperature effects influence
the populations mainly in the low-density region. 
In fact, the presence of tails in the Fermi distribution makes it possible 
to create particles at any density and thus typical production thresholds,
like for muon creation, disappear at finite temperature.
On the other hand, thermal effects are less important at high density, 
where the particle populations do not change appreciably when 
increasing temperature. 

The same considerations hold true when hyperons are included 
(lower panels).
In this case, we notice that the $\sgm$ and $\la$ thresholds disappear 
and that hyperons become more and more abundant at low density with 
increasing temperature.
The lepton fractions at low density amount to about 10$\%$ 
of the total particle population if the temperature is high. 
We notice that the appearance of the $\sgm$ induces a rapid 
deleptonization, no matter the value of the temperature. 
This happens because it is energetically more convenient to maintain the
charge neutrality through $\sgm$ formation than $\beta$ decays. 
The stellar core turns out to be mainly populated by neutrons and protons,
with an appreciable fraction ($\approx 30\%$) of hyperons. 

In Fig.~\ref{f:xinu} we show the particle fractions as a function of baryon 
density, for different values of the temperature, 
in the case that neutrinos are trapped. 
As in the previous figure, 
the upper panels show the particle fractions when stellar matter 
contains only nucleonic species, whereas the lower panels include 
the appearance of noninteracting $\sgm$ and $\la$.
The electron fraction is larger in neutrino-trapped 
matter than in the neutrino-free case; 
therefore the proton population is larger as well, 
and altogether the resulting EOS will be softer.
The appearance of muons is shifted to higher density values, 
because the onset is now determined by the difference between the 
electron and the neutrino chemical potentials, $\mu_e - \mu_{\nu_e}$.
However, if the temperature increases, muons appear at lower density.

Neutrino trapping has also strong consequences for the onset of hyperons.
In fact, the onset of the $\sgm$ is shifted to high density, whereas 
$\la$'s appear at slightly smaller density. 
This is due to the fact that the $\sgm$ onset depends on
the neutron and lepton chemical potentials, i.e., 
$\mu_n + \mu_e - \mu_{\nu_e}$, which stays at 
larger values in neutrino-trapped matter than in the neutrino-free case, 
thus delaying the appearance of the $\sgm$ to higher baryon density
and limiting its population to a few percent. 
On the other hand, the $\la$ onset depends on the neutron chemical 
potential only, which keeps at lower values in the neutrino-trapped case. 
We notice that the appearance of the $\sgm$ hinders 
muon production, which are not present at all.
When the temperature increases, more and more hyperons 
are present also at low densities, but they still remain a tiny fraction
of the total baryon density in this region of the protoneutron star.
Altogether, the hyperon fractions are much smaller than in the neutrino-free
matter.

Fig.~\ref{f:xiint} shows for comparison the particle fractions 
with interacting hyperons in cold beta-stable matter with and without
neutrinos.
With respect to the previous results assuming noninteracting hyperons, 
one observes in neutrino-free matter a shift of the $\la$ threshold
to substantially lower density, 
whereas the $\sgm$ onset remains practically fixed. 
This can be related to the properties of the hyperon single-particle
potentials at the relevant densities (Baldo et al. 1998, 2000a).
In neutrino-rich matter, the repulsive nature of the interactions at high 
density leads now even to a complete suppression of the $\sgm$,
and only the $\la$ is present in the matter.

\subsection{Equation of state}

\begin{figure*}[t] %...........................................................
\centering
\includegraphics[height=17cm,angle=270,bb=70 10 570 690,clip]{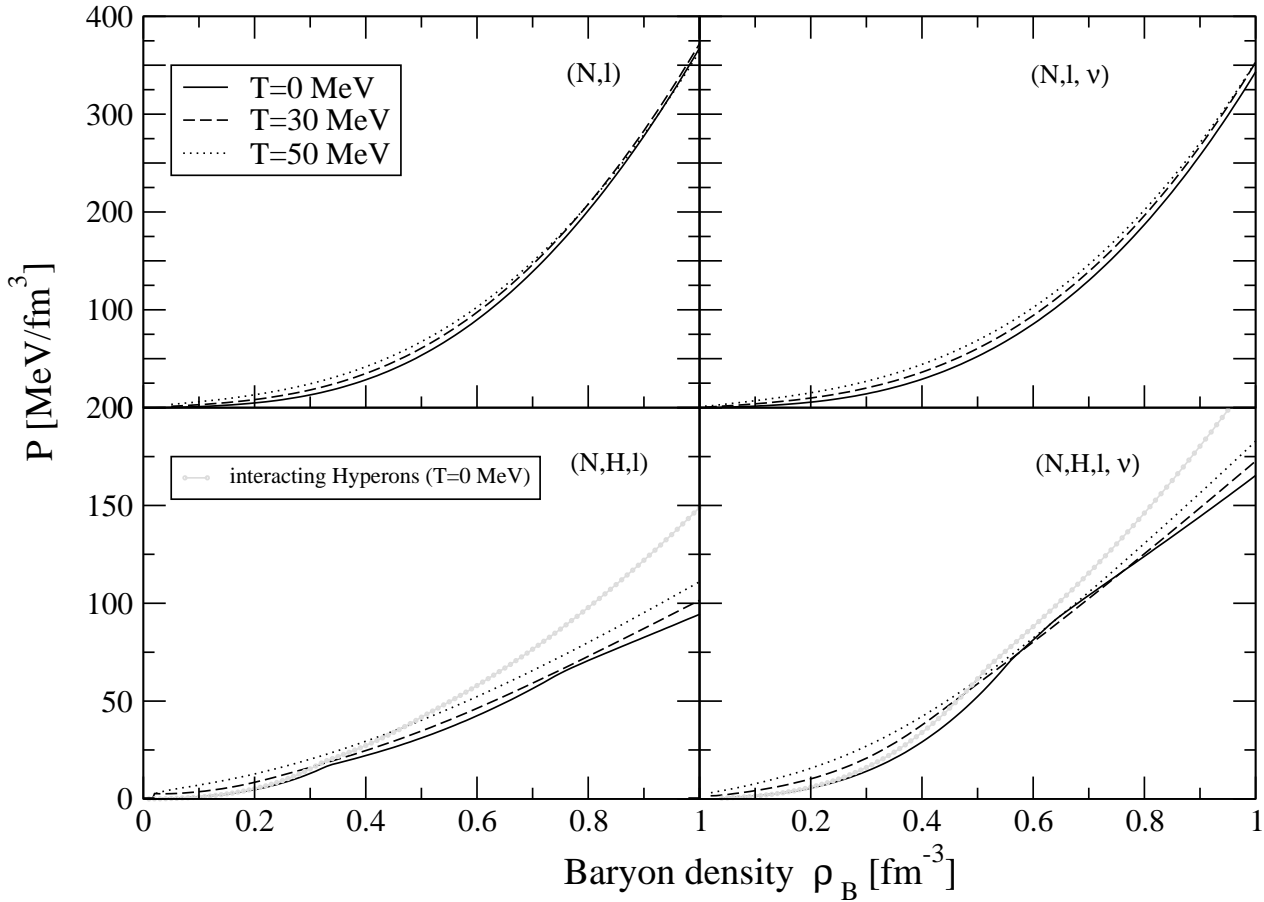}
\caption{
Pressure as a function of baryon density for 
beta-equilibrated matter at temperatures $T=0$, 30, and 50 MeV. 
The left-hand (right-hand) panels show the EOS in 
neutrino-free (neutrino-trapped) matter, 
with nucleons only (upper panels) and nucleons plus hyperons (lower panels).
The lower panels show results with interacting hyperons at zero temperature
(faint lines) and with free hyperons at different temperatures.}
\label{f:eos}
\end{figure*} %................................................................

\begin{figure*}[t] %...........................................................
\includegraphics[width=11.5cm,angle=270,bb=50 30 480 30]{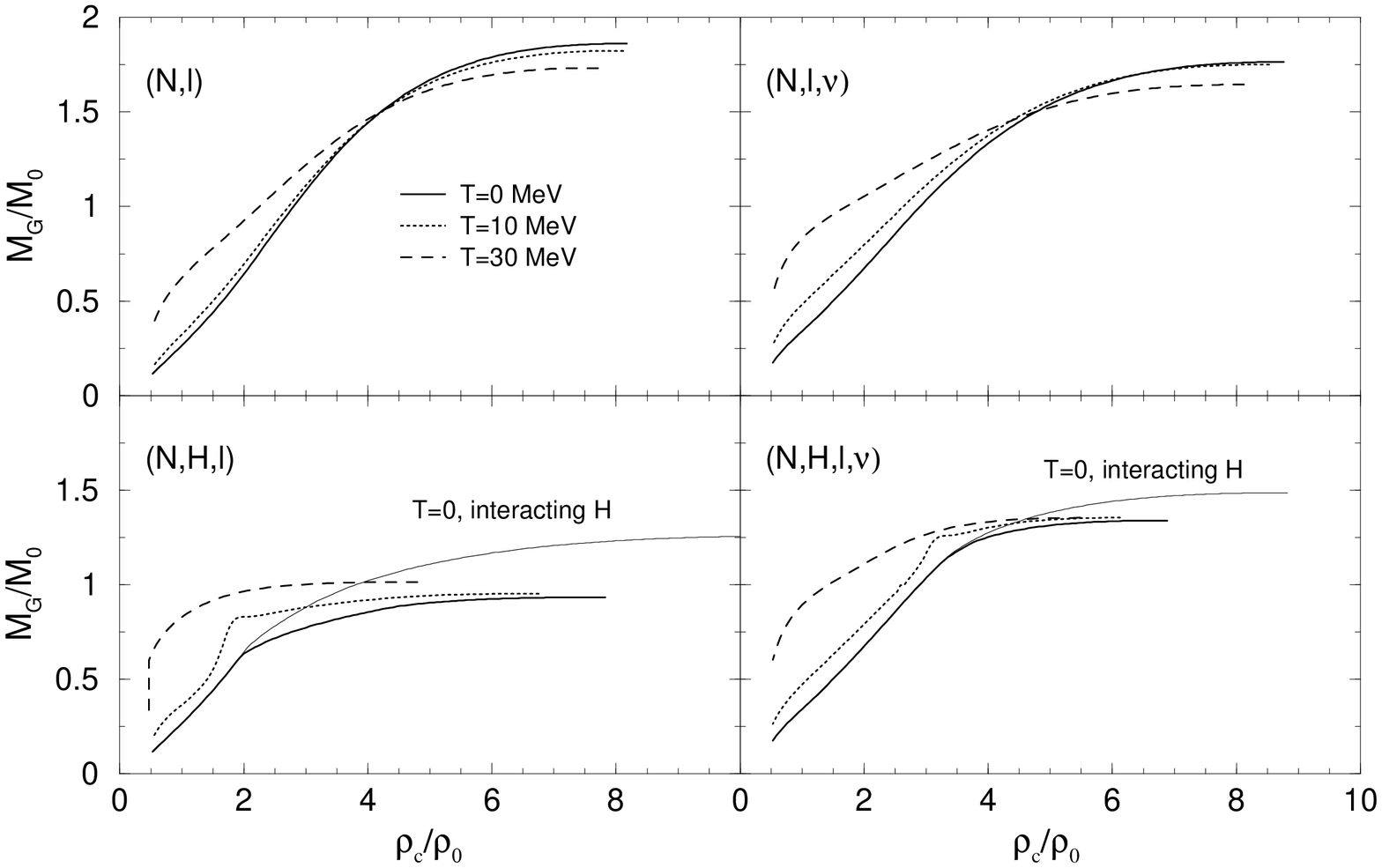}
\caption{
The gravitational mass 
(in units of the solar mass $M_\odot=1.98\times 10^{33}$g) 
as a function of the central baryon density 
(normalized with respect to the saturation value 
$\rho_0=0.17\;\mathrm{fm}^{-3}$) 
at temperatures $T=0$, 30, and 50 MeV. 
The left-hand (right-hand) plots regard neutrino-free (neutrino-trapped) matter.
The upper plots describe configurations without hyperons, 
while in the lower plots
the thin solid curves denote configurations of cold stars employing
interacting hyperons, whereas the remaining curves show warm stars at different
temperatures using noninteracting hyperons.}
\label{f:m}
\end{figure*} %................................................................

The resulting EOS is displayed in Fig.~\ref{f:eos}, 
where the pressure for beta-stable asymmetric matter, 
with and without neutrinos, 
is plotted as a function of the baryon density
at temperatures $T=0$, 30, and 50 MeV.
Let us begin with discussing the case of neutrino-free matter, 
shown without (upper left plot) and with hyperons (lower left plot). 
We notice that thermal effects produce a slightly stiffer EOS
with respect to the cold case, and that at very high densities 
they almost play no role. 
The inclusion of hyperons produces a dramatic effect, 
because the EOS gets much softer 
(slightly less with interacting hyperons due to the repulsive effect of
interactions at high density),
no matter the value of the temperature. 
We observe, for the $T=0$ case, 
two wiggles associated with the onset of the $\sgm$ and $\la$ hyperons, 
which however tend to disappear when temperature is introduced, 
due to the disappearance of the thresholds, as shown in Fig.~\ref{f:xi}. 

In the right-side panels we show the corresponding neutrino-trapped case. 
The EOS is slightly softer than in the neutrino-free case if only nucleons 
and leptons are present in the stellar matter. 
Again, the presence of hyperons introduces a strong softening of the EOS,
but less than in the neutrino-free case, because now the hyperons appear later
in the matter and their concentration is lower.  
A single large wiggle is now present at $T=0$ 
in the density range where both $\sgm$ and $\la$ appear. 
Thermal effects are rather small also in this case, 
except for the disappearance of the hyperon onsets.

\subsection{Global stellar structure}

In Fig.~\ref{f:m} we show the gravitational mass 
(in units of the solar mass $M_\odot$) 
as a function of the central baryon density
for stars without (upper panels) and with (lower panels) hyperons.
In the former case we notice that both finite temperature and neutrino trapping
reduce slightly (by about $0.1\;M_\odot$)
the value of the maximum mass, which is about $1.8\;M_\odot$
for a cold neutron star. 
This is due to the fact that at high baryon density 
the EOS gets softer with increasing temperature. 
This behavior is at variance with the results reported in 
Prakash et al. (1997),
where the critical mass increases with increasing entropy and,
therefore, also with increasing temperature.

In this regard, we should notice that in our calculations we are 
considering an isothermal profile, whereas in Prakash et al. (1997) 
the profile is isentropic. 
However, this cannot be the origin of the observed difference in the 
change of the maximum mass as a function of temperature and entropy. 
In fact, the protoneutron star stability is
mainly governed by the EOS in the central high-density part, and
therefore an increase of entropy or temperature should be equivalent.
Thus the value of the maximum mass at a given temperature in an
isothermal calculation could be traced back to the corresponding isentropic
calculation just from the entropy-temperature relationship for the
EOS at the high density present in the core of the star.
A more likely explanation can be found in the behavior of the EOS
with increasing temperature in our microscopic calculations. 
In fact, the decrease of maximum mass with temperature is in our case 
due to the decrease of asymmetry and to the consequent softening of the EOS.
It has to be noticed that for a completely local interaction there is no
dependence on temperature of the EOS except for the kinetic part.
In Prakash et al. (1997) the interaction is indeed mostly local, with only a
non-local correction, whose contribution is unfortunately not separately
reported. 
In our calculations the whole interaction part is temperature-dependent, 
due to the intrinsic non-locality of the single-particle potentials, 
and therefore the temperature dependence can be quite
different and larger.

The lower panels of Fig.~\ref{f:m} show the configurations of stars 
containing hyperons.
We observed in Fig.~\ref{f:eos} that the EOS softens considerably when 
hyperons are included, both in neutrino-free and neutrino-trapped matter.  
As a consequence the mass -- central density relation is also significantly 
altered and the value of the critical mass decreases by a large amount,
down to about $1.3\;M_\odot$ for neutron stars and $1.5\;M_\odot$ 
for protostars.
Since the former value falls below the mass of the best observed pulsar, 
i.e., $1.44~M_\odot$,
the EOS of high-density nuclear matter comprising only baryons 
(nucleons and hyperons) is probably unrealistic 
(even taking into account the present uncertainty
of hyperonic two-body and three-body forces)
and must be supplemented by a transition to quark matter.
This has been discussed extensively in 
(Burgio et al. 2002a, 2002b; Baldo et al. 2003; Maieron et al. 2004). 

Nevertheless, it is interesting to observe a behavior opposite to those
of nucleonic stars, which could even survive the inclusion of quark matter,
namely the maximum mass of hyperonic protostars is larger 
(by about $0.3\;M_\odot$)
than the one of cold neutron stars.
The reason is the minor importance of hyperons in the neutrino-saturated matter,
which leads to a stiffer equation of state, see Fig.~\ref{f:eos},
and to a larger maximum mass.
This feature could lead to metastable stars suffering 
a delayed collapse while cooling down, 
as discussed in (Prakash et al. 1997; Pons et al. 1999).
Our main results for the maximum mass configurations are summarized in
Table~\ref{t:max}.

\begin{table} %................................................................
%\squeezetable
\caption{
 Characteristics of the maximum mass configurations for different
 stellar compositions and temperatures.}
\bigskip
%\begin{ruledtabular}
\begin{tabular}{l|c|rrrrr}
 Composition           & $T$ (MeV) & $M/M_\odot$ & $R$ (km) & $\rho_c/\rho_0$\\
\hline
                       & 0         & 1.86        & 9.5      & 8.2            \\
 $N,l$                 & 10        & 1.82        & 9.5      & 8.1            \\
                       & 30        & 1.73        & 9.7      & 7.7            \\
\hline
                       & 0         & 1.76        & 9.1      & 8.8            \\
 $N,l,\nu$             & 10        & 1.75        & 9.2      & 8.5            \\
                       & 30        & 1.65        & 9.5      & 8.1            \\
\hline
                       & 0         & 0.93        & 10.2     & 7.8            \\
 $N,H_\mathrm{free},l$ & 10        & 0.95        & 11.0     & 6.7            \\
                       & 30        & 1.00        & 13.0     & 4.8            \\
\hline
 $N,H_\mathrm{int.},l$ & 0         & 1.25        &  8.8     & 11.5           \\
\hline
                       & 0         & 1.34        & 10.6     & 6.9            \\
 $N,H_\mathrm{free},l,\nu$& 10     & 1.35        & 11.0     & 6.1            \\
                       & 30        & 1.35        & 12.2     & 5.5            \\
\hline
 $N,H_\mathrm{int.},l,\nu$& 0      & 1.48        & 9.6      & 8.8            \\
\hline
\end{tabular}
%\end{ruledtabular}
%\footnotetext[1]{}
\label{t:max}
\end{table} %..................................................................

%------------------------------------------------------------------------------
\section{Conclusions}

In this paper we have studied the structure of (proto)neutron stars
on the basis of a microscopically derived EOS for baryonic matter
at finite temperature. 
Configurations with or without trapped neutrinos were considered. 
It was found that for nucleonic stars the thermal effects 
systematically reduce the maximum masses with increasing temperature. 
This effect is argued to be related to the strong non-locality of the 
mean fields as obtained from the microscopic calculations. 
Neutrino trapping further reduces slightly the maximum mass, 
as a consequence of the additional softening of the EOS, 
but the dominant effect appears to be produced by the temperature, 
if values up to 30 MeV are considered. 
These results indicate that the maximum mass of a nucleonic neutron star 
could be determined not by its mechanical stability but rather by the stability
of the hot protoneutron star progenitor.

On the contrary, 
we find that the maximum mass of a hyperonic protostar is substantially 
larger than the one of the cold star,
because both neutrino trapping and finite temperature tend to stiffen the EOS. 
Trapping shifts the onset of hyperons to considerably higher density
and reduces their concentrations, 
in particular the $\sgm$ hyperon appears only in the higher density region 
or disappears completely. 

However, as in the case of cold neutron stars, the addition of hyperons 
demands for the inclusion of quark degrees of freedom in order to obtain 
a maximum mass larger than the observational lower limit.
A consistent treatment of this aspect is left to a future work.

%------------------------------------------------------------------------------
%\begin{acknowledgements}
%\end{acknowledgements}

%------------------------------------------------------------------------------
\section*{Appendix A. Parametrizations of internal and free energy}

\begin{table} %................................................................
%\squeezetable
\caption{
 Parameters of the EOS fits, Eqs.~(\ref{e:fite},\ref{e:fitf}), 
 for symmetric nuclear matter (SNM) and pure neutron matter (PNM).}
\bigskip
%\begin{ruledtabular}
\begin{tabular}{l|rrrccccc}
                    &$a_1$ &$a_2$ &$b_0$ &$b_1$ &$b_2$ &$c_0$ &$c_1$ &$d$ \\
\hline
 $\!\!\!\!E/A$, SNM & 105  & 74   & -473 & -464 &      & 586  & 381  & 1.26 \\
 $\!\!\!\!E/A$, PNM & 109  & 64   &   34 & -240 &      & 249  & 164  & 1.97 \\
 $\!\!\!\!F/A$, SNM & 41   & 116  & -180 &      & -174 & 293  &      & 1.57 \\
 $\!\!\!\!F/A$, PNM & 21   & 116  &  101 &      & -131 & 191  &      & 2.62 \\
\end{tabular}
%\end{ruledtabular}
%\footnotetext[1]{}
\label{t:fit}
\end{table} %..................................................................

For the determination of the composition of beta-stable matter and
for the solution of the TOV equations,
it is very useful to provide analytical fits of the internal energy
$E/A(T,\rho,x_p)$ as well as the free energy $F/A(T,\rho,x_p)$.

It turns out that for both quantities the dependence on proton fraction 
can be very well approximated by a quadratic dependence, as
at zero temperature (Bombaci \& Lombardo 1991; Baldo et al. 1998, 2000a):
\be
 {E\over A}(T,\rho,x) \approx 
 {E\over A}(T,\rho,x=0.5) + (1-2x)^2 E_\mathrm{sym}(T,\rho) \:,
\label{e:parab}
\ee
where
the symmetry energy $E_\mathrm{sym}$ can be expressed in
terms of the difference of the energy per particle between pure neutron 
($x=0$) and symmetric ($x=0.5$) matter:
\bea
  E_\mathrm{sym}(T,\rho) &=& 
  - {1\over 4} {\partial(E/A) \over \partial x}(T,\rho,0)
\\
  &\approx& {E\over A}(T,\rho,0) - {E\over A}(T,\rho,0.5) \:.
\label{e:sym}
\eea
Therefore, it is only necessary to provide parametrizations
of both quantities for symmetric nuclear matter
and pure neutron matter.
We find that the following functional forms provide excellent parametrizations
of the numerical results:
\bea
 {E\over A}(T,\rho) &=& (a_1t+a_2t^2) + (b_0+b_1t)\rho + (c_0+c_1)\rho^d \:,
\label{e:fite}
\\
 {F\over A}(T,\rho) &=& (a_1t+a_2t^2)\ln(\rho) + (b_0+b_2t^2)\rho + c_0\rho^d \:,
\label{e:fitf}
\eea
where $t=T/(100\;\mathrm{MeV})$ and $E,F$ and $\rho$ are given in
MeV and fm$^{-3}$, respectively.
The parameters of the different fits are given in Table~\ref{t:fit}.

%------------------------------------------------------------------------------
%\section*{Appendix B. Thermodynamics of noninteracting particles ??}

%------------------------------------------------------------------------------


\begin{thebibliography}{}

\bibitem[1997]{pan1} 
Akmal, A., \& Pandharipande, V. R. 1997, 
Phys. Rev. C, 56, 2261 

\bibitem[1998]{pan2}  
Akmal, A., Pandharipande, V. R., \& Ravenhall, D. G. 1998,
Phys. Rev.  C, 58, 1804

\bibitem{bbb}
Baldo, M., Bombaci, I. \& Burgio, G. F. 1997, 
A\&A, 328, 274 

\bibitem{bbs1}
Baldo, M., Burgio, G. F., \& Schulze, H.-J. 1998, 
Phys. Rev. C, 58, 3688 \\
------. 2000a, Phys. Rev. C, 61, 055801 

\bibitem{book}
Baldo, M. 1999, 
Nuclear Methods and the Nuclear Equation of State,
ed. World Scientific, Singapore 

\bibitem{big1} 
Baldo, M., \& Ferreira, L. S. 1999, 
Phys. Rev. C, 59, 682 

\bibitem{th1}
Baldo, M., Song, H. Q., Giansiracusa, G., \& Lombardo, U. 2000b,
Phys. Lett. B, 473, 1 

\bibitem{th2}
Baldo, M., Fiasconaro, A., Song, H. Q., Giansiracusa, G., \& Lombardo, U. 2001,
Phys. Rev. C, 65, 017303

\bibitem{njl}
Baldo, M., Buballa, M., Burgio, G. F., Neumann, F., Oertel, M., 
\& Schulze, H.-J. 2003,
Phys. Lett. B, 562, 153

\bibitem{big2} 
Baldo, M., Ferreira, L. S., \& Nicotra, O. E. 2004,  
Phys. Rev. C, 69, 034321

\bibitem{baym}
Baym, G., Pethick, C., \& Sutherland, D. 1971,
ApJ, 170, 299 

\bibitem{bloch}
Bloch, C., \& De Dominicis, C. 1958, 
Nucl. Phys., 7, 459 \\
------. 1959a, Nucl. Phys., 10, 181 \\
------. 1959b, Nucl. Phys., 10, 509
 
\bibitem{bom}
Bombaci, I., \& Lombardo, U. 1991,
Phys. Rev. C, 44, 1892

\bibitem{bon} 
Bonche, P., Chabanat, E., Haensel, P., Meyer, J., \& Schaeffer, R. 1998, 
Nucl. Phys. A, 643, 441 

\bibitem{bbsss} 
Burgio, G. F., Baldo, M., Sahu, P. K., Santra, A. B., \& Schulze, H.-J. 2002a,
Phys. Lett. B, 526, 19 

\bibitem{bbss}
Burgio, G. F., Baldo, M., Sahu, P. K., \& Schulze, H.-J. 2002b, 
Phys. Rev. C, 66, 025802 

\bibitem{burr}
Burrows, A., \& Lattimer, J. M. 1986, 
ApJ, 307, 178

\bibitem{carl}
Carlson, J.,  Pandharipande, V. R., \& Wiringa, R. B. 1983, 
Nucl. Phys. A, 401, 59

\bibitem{cug00} 
Cugnon, J., Lejeune, A., \& Schulze, H.-J. 2000, 
Phys. Rev. C, 62, 064308

\bibitem{fey}
Feynman, R., Metropolis, F., \& Teller, E. 1949,
Phys. Rev., 75, 1561 

\bibitem{FriedPand} 
Friedman, B., \& Pandharipande, V. R. 1981,  
Nucl. Phys. A, 361, 502 

\bibitem{glen}
Glendenning, N. K. 1982, 
Phys. Lett. B, 114, 391 \\
------. 1985, ApJ, 293, 470 \\
------. 2000, Compact Stars, Nuclear Physics, Particle Physics, 
and General Relativity, 2nd ed., Springer-Verlag, New York

\bibitem{huber} 
Huber, H., Weber, F., \& Weigel, M. K. 1999, 
Phys. Rev. C, 57, 3484 

\bibitem{chi} 
Kaiser, N., Fritsch, S., \& Weise, W. 2002, 
Nucl. Phys.  A, 697, 255 

\bibitem{lac80}
Lacombe, M., Loiseau, B., Richard, J. M., Vinh Mau, R., C\^ot\'e, J., 
Pir\`es, P., \&  de Tourreil, R. 1980, 
Phys. Rev. C, 21, 861

\bibitem{lat}
Lattimer, J. M., Prakash, M., Masak, D., \& Yahil, A. 1990, 
ApJ, 355, 241

\bibitem{lej}
Lejeune, A., Grang\a'{e}, P., Martzolff, M., \& Cugnon, J. 1986,
Nucl. Phys. A, 453, 189

\bibitem{li}
Li, G. Q., Machleidt, R., \& Brockmann, R. 1992, 
Phys. Rev. C, 45, 2782 

\bibitem{dbhf}
Machleidt, R. 1989,  
Adv. Nucl. Phys., 19, 189 

\bibitem{mae89}
Maessen, P., Rijken, Th., \& de Swart, J. 1989, 
Phys. Rev. C, 40, 2226

\bibitem{mai}
Maieron, C., Baldo, M., Burgio, G. F., \& Schulze, H.-J. 2004,
Phys. Rev. D, 70, 043010

\bibitem{pan3} 
Morales, J., Pandharipande, V. R., \& Ravenhall, D. G. 2002,
Phys. Rev. C, 66, 054308

\bibitem{myers}
Myers, W. D., \& Swiatecki, W. J. 1996, 
Nucl. Phys. A, 601, 141

\bibitem{nv}
Negele, J. W., \& Vautherin, D. 1973,
Nucl. Phys. A, 207, 298 

\bibitem{pons}
% EVOLUTION OF PROTO-NEUTRON STARS
Pons, J. A., Reddy, S., Prakash, M., Lattimer, J. M., \& Miralles, J. A. 1999, 
ApJ, 513, 780

\bibitem{pra}
Prakash, M., Bombaci, I., Prakash, M., Ellis, P. J., 
Lattimer, J. M., \& Knorren, R. 1997, 
Phys. Rep., 280, 1

\bibitem{pand}
Schiavilla, R., Pandharipande, V. R., \&  Wiringa, R. B. 1986, 
Nucl. Phys. A, 449, 219 

\bibitem{hypmat1} 
Schulze, H.-J., Lejeune, A., Cugnon, J., Baldo, M., \& Lombardo, U. 1995,
Phys. Lett. B, 355, 21 

\bibitem{hypmat2} 
Schulze, H.-J., Baldo, M., Lombardo, U., Cugnon, J., \& Lejeune, A. 1998,
Phys. Rev. C, 57, 704 

\bibitem{wal} 
Serot, B. D., \& Walecka, J. D. 1986, 
Adv. Nucl. Phys., 16, 1

\bibitem{shapiro}
Shapiro, S. L., \& Teukolsky, S. A. 1983, 
Black Holes, White Dwarfs, and Neutron Stars, 
ed. John Wiley \& Sons, New York

\bibitem{son} 
Song, H. Q., Baldo, M., Giansiracusa, G., \& Lombardo, U. 1998,
Phys. Rev. Lett., 81, 1584 

\bibitem{tayl}
Taylor, J. H., \& Weisberg, J. M. 1989, 
ApJ, 345, 434

\bibitem{Malfliet} 
Ter Haar, B., \& Malfliet, R. 1986, 
Phys. Rev. Lett., 56, 1237 \\
------. 1987, Phys. Rep., 149, 207 

\bibitem{barc} 
Vida\~na, I., Polls, A., Ramos, A., Engvik, L., \& Hjorth-Jensen, M. 2000, 
Phys. Rev. C, 62, 035801 

\bibitem{vid01} 
Vida\~na, I., Polls, A., Ramos, A., \& Schulze, H.-J. 2001,
Phys. Rev. C, 64, 044301

\bibitem{weber}
Weber, F. 1999, 
Pulsars as Astrophysical Laboratories for Nuclear and Particle Physics, 
ed. Institute of Physics Publishing, Bristol and Philadelphia

\bibitem{wiringa}
Wiringa, R. B., Stoks, V. G. J., \& Schiavilla, R. 1995, 
Phys. Rev. C, 51, 38

\bibitem{zhou}
Zhou, X. R., Burgio, G. F., Lombardo, U., Schulze, H.-J., \& Zuo, W. 2004, 
Phys. Rev. C, 69, 018801


\end{thebibliography}
\end{document}